\documentstyle[nocenter,sfbold,noindent]{thesis}
\begin{document}

\title{Kinetic equation for strongly interacting dense Fermi systems}
\author{
Pavel Lipavsk{\'y}$^1$, Klaus Morawetz$^2$ and V\'aclav \v Spi\v cka$^1$}
\institution{$^1$Institute of Physics, Academy of Sciences, Cukrovarnick\'a 10,
16200 Praha 6, Czech Republic\\
$^2$Max-Planck-Institute for the Physics of Complex Systems, 
Noethnitzer Str. 38, 01187 Dresden, Germany}
\date{}
\maketitle
%

\begin{abstract} 
We review the non-relativistic Green's-function approach to the kinetic 
equations for Fermi liquids far from equilibrium. The emphasis is on 
the consistent treatment of the off-shell motion between collisions
and on the non-instant and non-local picture of binary collisions.

The resulting kinetic equation is of the Boltzmann type, and it
represents
an interpolation between the theory of transport in metals and the 
theory of moderately dense gases. The free motion of particles is
renormalised by various mean field and mass corrections in the spirit
of Landau's quasiparticles in metals. The collisions are non-local in 
the spirit of Enskog's theory of non-ideal gases. The collisions are
moreover non-instant, a feature which is absent in the theory of gases,
but which is
shown to be important for dense Fermi systems. 

In spite of its formal complexity, the presented theory has a simple
implementation within the Monte-Carlo simulation schemes. Applications 
in nuclear physics are given for heavy-ion reactions and the results are
compared with the former theory and recent experimental data.

The effect of the off-shell motion and the non-local and non-instant
collisions on the dynamics of the system can be characterised in 
terms of thermodynamic functions such as the energy density or 
the pressure tensor. Non-equilibrium counterparts of these functions 
and the corresponding balance equations are derived and discussed from
two points of view. Firstly, they are used to prove the conservation
laws. 
Secondly, the role of individual microscopic mechanisms in fluxes of
particles and momenta and in transformations of the energy is clarified.

\newpage

Nous examinons la technique des fonctions de Green non relativistes
appliqu\'ee aux \'equations cin\'etiques pour les liquides de Fermi hors
\'equilibre. L'accent est mis sur le traitement coh\'erent des effets 
hors couche  entre les collisions ainsi que sur l'aspect non
local et non instantan\'e des collisions binaires.\\

L'\'equation cin\'etique r\'esultante est de type Boltzmann et repr\'esente
une interpolation entre la th\'eorie du transport dans les m\'etaux et la
th\'eorie des gaz mod\'er\'ement denses. Le mouvememt libre des particules
est renormalis\'e par diverses corrections de masse et de champ moyen dans
le m\^eme esprit que pour les quasi-particules de Landau dans les m\'etaux.
Les collisions sont
non locales au sens de la th\'eorie d'Enskog des gaz r\'eels. De plus ces
collisions ne sont pas instantan\'ees, caract\'eristique absente de la
th\'eorie des gaz, mais dont nous montrons l'importance dans les syst\`emes de
Fermi denses.\\

Malgr\'e sa complexit\'e formelle, la th\'eorie que nous pr\'esentons est
facile \`a implanter dans  les simulations Monte-Carlo. Nous appliquons
notre m\'ethode aux r\'eactions d'ions lourds en physique nucl\'eaire et
confrontons les r\'esultats \`a ceux de l'ancienne th\'eorie ainsi qu'aux
donn\'ees exp\'erimentales r\'ecentes.\\

Les effets  hors  couche, de la non localit\'e et de la non
instantan\'eit\'e des collisions sur la dynamique du syst\`eme peuvent se
traduire en termes de fonctions thermodynamiques telles que la densit\'e
d'\'energie ou le tenseur de pression. Nous explicitons les \'equivalents hors
\'equilibre de ces fonctions ainsi que les \'equations bilans associ\'ees et
nous les discutons de deux points de vue diff\'erents: premi\`erement pour
prouver les lois de conservation et deuxi\`ement pour clarifier le r\^ole des
m\'ecanismes microscopiques individuels dans les flux de particules ou
d'impulsions et dans les transformations de l'\'energie.  
  
\end{abstract}

\chapter{Intention}

The initial intention was to prepare a review article which covers
the latest progress in a variety of physical fields dealing with 
non-equilibrium many-body systems. During the last decade, kinetic 
theories developed within the chemical, nuclear, plasma and 
solid-state physics have started to merge whereas previously they were
based 
on rather 
distinct approaches. It turned out that to cover all essential
concepts on a basis of a unified theory exceeds the frame of a review
article. We have been encouraged by a community of theoretical 
physicists to extend the manuscript into a monograph.

The manuscript is aimed for a wide audience of theoretical physicists 
interested in non-equilibrium many body systems. Selected chapters
can be studied by advanced undergraduate students. Advanced chapters
(denoted by stars) are addressed to a growing community of physicists
having at least rudimentary knowledge of Green's functions. We expect 
particular
interest from physicists involved in nuclear matter only for 
which field we give applications as not to drive reader's attention 
into many directions. For non-experts in nuclear matter, we first 
review properties of nuclear matter which are discussed in the light 
of experience from other fields.

The first part reviews selected work from kinetic theories of 
five different systems: moderately dense gases, electronic Fermi liquids 
in metals, electronic transport in semiconductors, non-ideal plasma, and
nuclear matter. These scattered topics are used to introduce and
enlighten physical properties studied later within the kinetic theory.

The second part provides the rigorous derivation from non-equilibrium 
quantum statistics. All steps are introduced in their historical 
background, from the simplest approximations to their recent form. 
This scheme allows the reader to benefit from preliminary 
knowledge of the kinetic theory (the rudimentary kinetic theory belongs 
to introductory courses for physicists) and to identify a suitable
point to start. 

The third part discusses the kinetic theory and its implementation to
nuclear physics. In the figures we present realistic values
of important physical quantities and demonstrate the accuracy of various 
approximations. This part also provides practical advice for
implementation of non-local corrections into numerical treatments.
They are explained in the context of quantum-molecular-dynamics simulations of 
heavy ion reactions.

The fourth part is devoted to thermodynamic properties implied by the
kinetic theory. It includes the law of acting masses, the particle 
flows, the density of energy, and the stress tensor. These properties
are discussed on two levels, from balance equations and from their
quantum statistical definitions. The first level, which explains a 
link between the kinetic theory and thermodynamics, allows the reader
to learn thermodynamic relations in an easy way. The second level is
aimed for experts in non-equilibrium quantum statistics. We prove 
the internal consistency of the kinetic theory and discuss shortcomings
of previous approaches.

The text is structured so that a reader not interested in technical
details can follow selected chapters. In this case, the reader will
approach the theory from the quasiclassical picture and learn how to
implement it in practise.

\chapter{Source}

The full text of the book is available at:

http://www.ed-phys.fr/articles/anphy/abs/2001/01/annales012001/annales012001.html

or can be ordered as book: ISBN: 2-86883-541-4

********************************************************

Klaus Morawetz

Max-Planck-Institut fuer Physik komplexer Systeme

Noethnitzer Str. 38, 01187 Dresden

Tel: +(49) 351 871 1126 Fax: +(49) 351 871 1999

e-mail: morawetz@mpipks-dresden.mpg.de

http://www.mpipks-dresden.mpg.de/~morawetz

********************************************************

\end{document}